\documentclass[sigconf,pdflatex]{acmart}%,dvipdfm
\AtBeginDocument{%
	}

\setcopyright{acmlicensed}
\copyrightyear{2025}
\acmYear{2025}
\acmDOI{1234567.1234567}
\acmConference[SIGMOD '25]{ACM SIGMOD International Conference on Management of Data}{June 22--27,
	2025}{Berlin, Germany}
\acmISBN{978-1-4503-1234-1/2025/06}

\acmSubmissionID{16}

\begin{document}
	
\title{LEDD: Large Language Model-Empowered Data Discovery\\in Data Lakes}

\author{Qi An}
\affiliation{%
	\institution{Tsinghua University}
	\city{Beijing}
	\country{China}
}

\author{Chihua Ying}
\affiliation{%
	\institution{Xiamen University}
	\city{Xiamen}
	\country{China}}

\author{Yuqing Zhu}
\authornote{zhuyuqing@tsinghua.edu.cn}
\affiliation{%
\institution{Tsinghua University}
\city{Beijing}
\country{China}}

\author{Yihao Xu}
\affiliation{%
	\institution{Tsinghua University}
	\city{Beijing}
	\country{China}}

\author{Manwei Zhang}
\affiliation{%
	\institution{Tsinghua University}
	\city{Beijing}
	\country{China}}

\author{Jianmin Wang}
\affiliation{%
	\institution{Tsinghua University}
	\city{Beijing}
	\country{China}}
\renewcommand{\shortauthors}{An et al.}

\begin{abstract}
Data discovery in data lakes with ever increasing datasets has long been recognized as a big challenge in the realm of data management, especially for semantic search of and hierarchical global catalog generation of tables. While large language models (LLMs) facilitate the processing of data semantics, challenges remain in architecting an end-to-end system that comprehensively exploits LLMs for the two semantics-related tasks. In this demo, we propose LEDD, an end-to-end system with an extensible architecture that leverages LLMs to provide hierarchical global catalogs with semantic meanings and semantic table search for data lakes. Specifically, LEDD can return semantically related tables based on natural-language specification. These features make LEDD an ideal foundation for downstream tasks such as model training and schema linking for text-to-SQL tasks. LEDD also provides a simple Python interface to facilitate the extension and the replacement of data discovery algorithms.
\end{abstract}



\maketitle
	
\section{Introduction}

Data discovery helps to locate tables and data in the large data lakes, which have continuously growing datasets coming from a wide range of sources. It is also necessary for the understanding of data value and content, as well as for downstream applications like model training. Researchers have built systems for data discovery~\cite{tableDiscovery}, supporting three pivotal tasks of keyword search, joinable table search, and unionable table search~\cite{lakeJoinTable,tableSearch24,tableSearchUnion}.

While hierarchical catalogs with semantic meanings are essential for the usability of data~\cite{googleGoods,annotateSemantic}, there is currently no automatic solution to a hierarchical global view of datasets in the data lake with semantic meanings. Existing approaches can only cluster datasets into scattered groups or apply semantic annotation~\cite{googleGoods,annotateSemantic}. The hierarchical global view aligns with users' way of understanding data. When combined with semantic queries, it can effectively facilitate users in data discovery. The core difficulty of hierarchical data organization lies in the understanding and processing of data semantics.

Large Language Models (LLMs) empowered data management is an emerging and important area~\cite{llmDB}. LLMs have shed light on tackling the complexities involved in semantic processing of data, e.g., question answering and data wrangling. While the semantic organization of global data is important and the LLM approach is promising, no system has yet been proposed that enables solving the former using the latter. Empowering data discovery with LLMs presents the following \emph{challenges}: i) how to effectively architect LLMs and data lakes to form a comprehensive solution; ii) how to tackle the complexities in aligning the semantic global view generation with the data/table search processing; and, iii) how to enable future extension of new algorithms.

To this end, we propose LEDD, an end-to-end prototype system to support LLM-empowered data discovery in data lakes. LEDD builds upon IGinX\footnote{https://iginx-thu.github.io/IGinX/}, a federated data lake system~\cite{federatedComputation} that supports flexible Python UDF (User-Defined Function) declarations to be used with SQL statements. LEDD extracts data from IGinX and implements processing algorithms in Python UDFs that exploit LLM services. Designed with a closed loop of processing workflows, LEDD integrates and enhances the data/table search processing with the semantic global view generation components. Building upon IGinX's Python UDF, the functions of LEDD can be easily extended and improved by replacing the corresponding UDF implementations.

At high level, LEDD features the following key functionalities:\\
\textbf{Hierarchical global data catalogs.} LEDD organizes all the tables and columns in the data lake into a hierarchical catalog with semantic meanings. Given a large set of scattered datasets, LEDD can cluster and summarize them iteratively and hierarchically until the final layer has members few enough to be grasped by users.\\
\textbf{Semantic table search and data discovery.} LEDD supports semantic search of tables and columns. Users can express their intent in natural language. LEDD locates all the relevant tables and columns, which are then organized into a hierarchical view with table and column names highlighted.\\
\textbf{Real-time analysis of column relations.} LEDD analyzes relations between nodes in the hierarchical graph view in real time during users' exploration of data. To facilitate exploration, LEDD first shows a collapsed view of the graph with only the highest level of summarized categories visible. When users click on a graph node, LEDD expands the node to show the sub-categories, displaying an extended graph whose new nodes are correlated with original nodes for important relations, e.g., joinable or unionable relations.\\
\textbf{Extensible algorithm interfaces.} LEDD supports the extension of algorithms in the form of IGinX UDFs, such that the algorithms for schema clustering, query embedding, and table/category relation analysis can be replaced or extended.

Overall, LEDD offers the following advantages. First, to the best of our knowledge, LEDD is the first prototype system that can automatically generate a hierarchical global data view with semantic meanings for data lakes, exploiting large language models and facilitating data discovery. Second, LEDD provides semantic search of tables, which is generally unsupported in previous data management systems or prototypes. Third, LEDD provides a comprehensive solution to the above functions while implementing a highly extensible architecture for users to extend new data discovery algorithms.

\begin{figure}[t]
	\centering
	\includegraphics[width=0.9\linewidth]{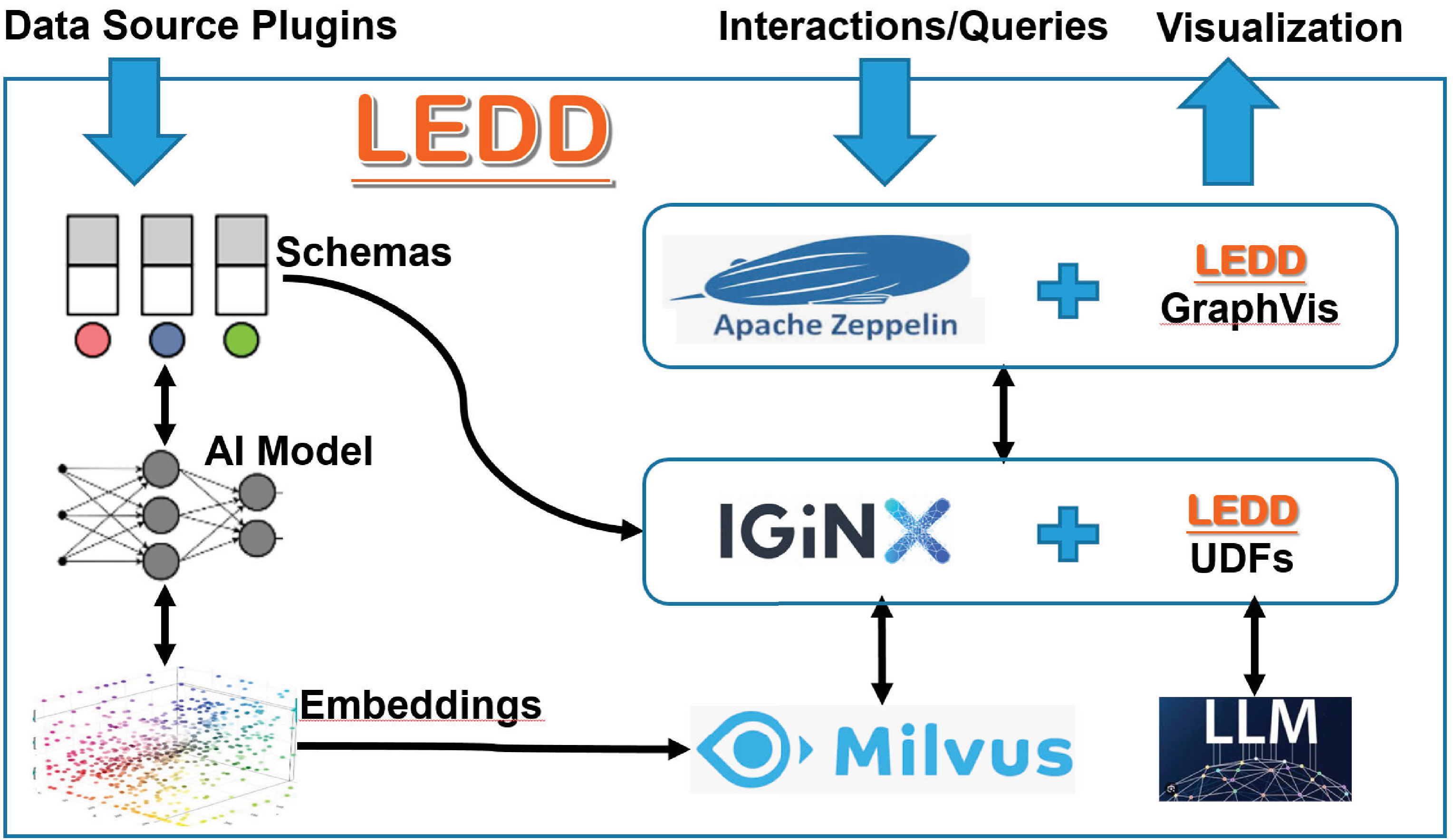}
	\caption{The system architecture of LEDD.}\vspace{-12pt}
	\label{fig:arch}
\end{figure}
\textbf{Existing Works.} Existing prototypes or systems primarily focus on the three tasks of table discovery, i.e., keyword search, joinable table search, and unionable table search~\cite{lakeJoinTable,tableSearch24,tableSearchUnion}. Aurum primarily supports keywords search of related tables~\cite{aurumStonebraker}, while LakeCompass is geared towards the integration of the three tasks for a data lake~\cite{LakeCompass}. InsightPilot leverages LLMs for data exploration, but its main focus is on recommending the next analytical actions rather than data discovery~\cite{InsightPilot}. Civilizer can only establish associations between existing columns and tables but cannot provide a global hierarchical view of data, which are essential for high-level abstraction and data comprehension~\cite{viewPolystore}. Google's Goods dataset management system automatically constructs a catalog and supports keyword queries along with some specific attribute queries, but it does not support semantic search of tables or hierarchical semantic view of data~\cite{googleGoods}. To sum up, hierarchical catalogs with semantic meanings and semantic search of tables are generally not considered comprehensively. LEDD employs a flexible and extensible system architecture to exploit a federated computation system IGinX and LLM, supporting both functions simultaneously.

In sum, LEDD is an end-to-end system with an extensible architecture that leverages LLMs to provide hierarchical global catalogs with semantic meanings and semantic table search for data lakes. 

\section{System Overview}

At a high level, LEDD (LLM-Empowered Data Discovery) provides three key functions: hierarchical global data catalogs with semantic meanings, semantic search of tables and columns, and real-time analysis of column relations. LEDD has been implemented in Python 3.11 and Javascript, leveraging OpenAI's API for LLM interaction, IGinX for data access and Zeppelin for basic visualization interface. LEDD will be open-sourced on Github.

\subsection{LEDD's Architecture}

The core functions of LEDD are implemented as an extended graph visualization component of Apache Zeppelin\footnote{https://zeppelin.apache.org/} and a set of UDFs in IGinX. Users can interact with LEDD through Zeppelin's web interface, using IGinX queries, LEDD commands, or graphical interactions. 

Data sources can be plugged into IGinX with schema or metadata automatically extracted. Building on the IGinX data lake, LEDD accesses global data and metadata within IGinX Python UDFs using the Python APIs. LEDD exploits AI models to generate semantic embeddings of tables and columns, as well as higher levels of categories. Instead of using LSH or HNSW indices~\cite{LakeCompass} like other data discovery systems, LEDD stores embeddings in the vector database Milvus\footnote{https://milvus.io/}, which offers efficient search of embedding vectors. LEDD's interactions with LLMs are also implemented within the UDFs. While LEDD has predefined LLM prompts for category summary and relation analysis, it exploits OpenAI's APIs for LLM access so that different LLMs can be used when necessary.

\begin{figure}[t]
	\centering
	\includegraphics[width=1.1\linewidth]{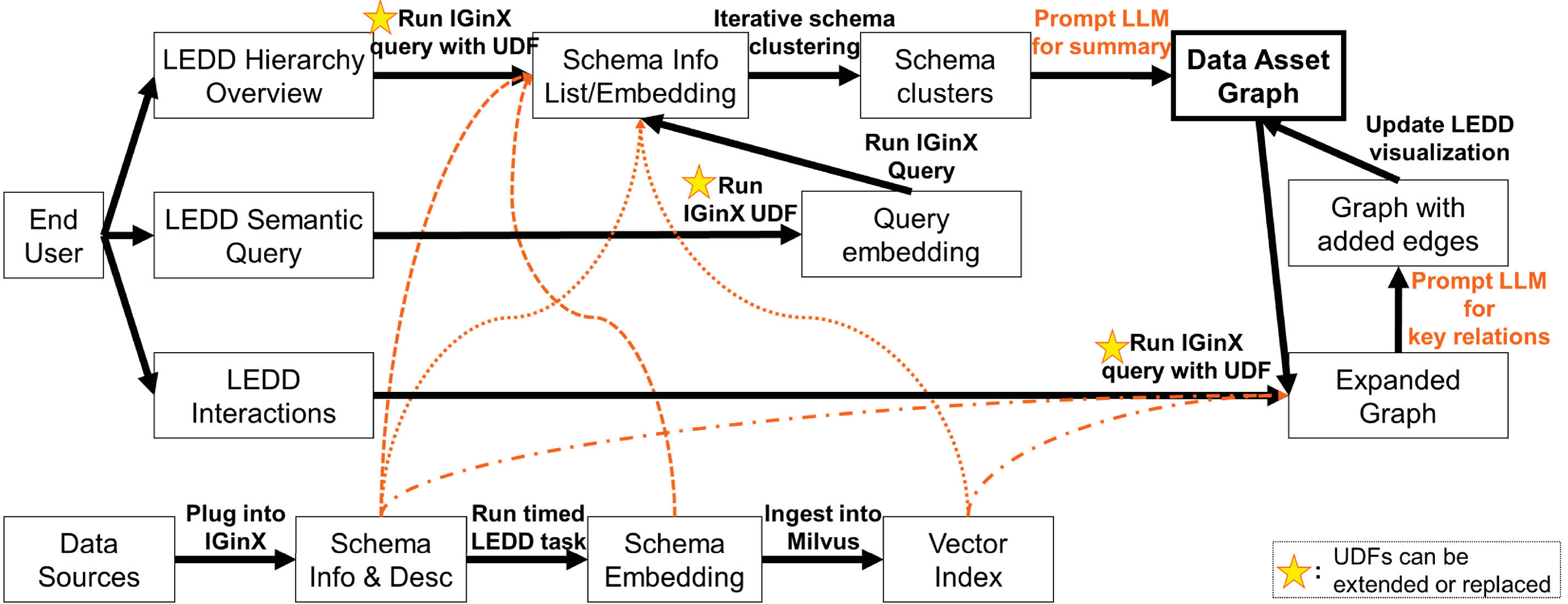}
	\caption{The processing workflows of LEDD's three key functions.}\vspace{-12pt}
	\label{fig:flow}
\end{figure}

In the following, we describe LEDD's three key functions, as well as its extensible interface for new algorithms. Their corresponding processing workflows are illustrated in Figure~\ref{fig:flow}. 

\subsection{Hierarchical Global Catalog Generation}
\label{sec:hierarchy}

Hierarchical global data catalogs with semantic meanings is essential for the usability of data. They represent the beginning of data exploitation. However, automatically constructing a hierarchical data catalog for an aggregated data lake has not been fully resolved due to (1) the difficulty in automatically identifying equivalence relationships between data schema, (2) challenges in semantically organizing hierarchical relationships between data, and (3) the inability of existing systems to directly integrate data without manual mapping while preserving information about the data sources. LEDD constructs a comprehensive solution, laying the foundation over IGinX and LLM to tackle the above challenges.

LEDD exploits IGinX to obtain the physical hierarchical relationships of data sources, their tables and the corresponding columns. Extracting from IGinX, LEDD integrates the following six facets of information to generate embeddings: (1) data source information, (2) the source-table-column path, (3) sibling columns, (4) value characteristics of the column, (5) related tasks the column has been involved in, and (6) description of the column (if available).

There are two approaches to utilizing column information, and LEDD adopts the latter: i) embedding each of the six facets into a vector, then integrating them into a unified vector through a fusion algorithm (i.e., composite embedding); and, ii) summarizing the six-dimensional information using LLM, followed by semantic embedding to obtain a unified representation vector. The LLM-based approach aligns more closely with human cognitive process for integrating knowledge and information. The reason the latter approach has not been used in the past is that such summarization capabilities became feasible only after the advent of LLMs.

LEDD performs clustering based on the resulting embedding vectors. Given a set of data source embeddings, each is first represented as a node in a graph. LEDD iteratively abstracts the corresponding graph into a super graph by clustering nodes to a given number $K$, which is set to the graph size best suited for visualization. LEDD allows different clustering algorithms to be used. For each iteration, LEDD asks the LLM to summarize each cluster of nodes into a category described by a short phrase. When the number of graph nodes reaches $K$, LEDD stops the clustering and summarizing process and displays the highest layer of categories in the hierarchy.
\subsection{Semantic Search and Relation Analysis}

In the visualization of the hierarchical catalog, LEDD supports semantic search of tables, as well as real-time relation analysis of newly expanded graph nodes. Given a query in natural language, LEDD calls the backend LLM to rephrase the query according to an example query and then calls an AI model for semantic embedding. Both the LLM and the AI model can be flexibly changed to any appropriate ones. The embedding vector is then searched against the index in Milvus, retrieving $n$ closest related vectors. If the search is initiated for a visualized graph, the returned vectors are retrieved from IGinX and highlighted in the graph. If the search is initiated from an empty view, all the retrieved tables and columns are organized into a new hierarchical graph, with disconnected nodes aggregated into layers of categories based on the aforementioned clustering method in Section~\ref{sec:hierarchy}.

LEDD dynamically generates relationships between nodes newly expanded from the graph and the original nodes during users' interactions with the visualized graph. Each newly added node is compared for similarity with nodes in the original graph, and the $m$ most similar nodes are queried for relationships using LLM. For closely related nodes, LLM generates relationship descriptions, which are then displayed on the interface. The thickness of the edges represents the degree of similarity between nodes, allowing users to easily identify the most similar nodes, facilitating the identification of tables or columns that can be joined or unioned with.\vspace{-6pt}

\subsection{Extensible Algorithm Interface}

\begin{figure}[t]
	\centering
	\includegraphics[width=\linewidth]{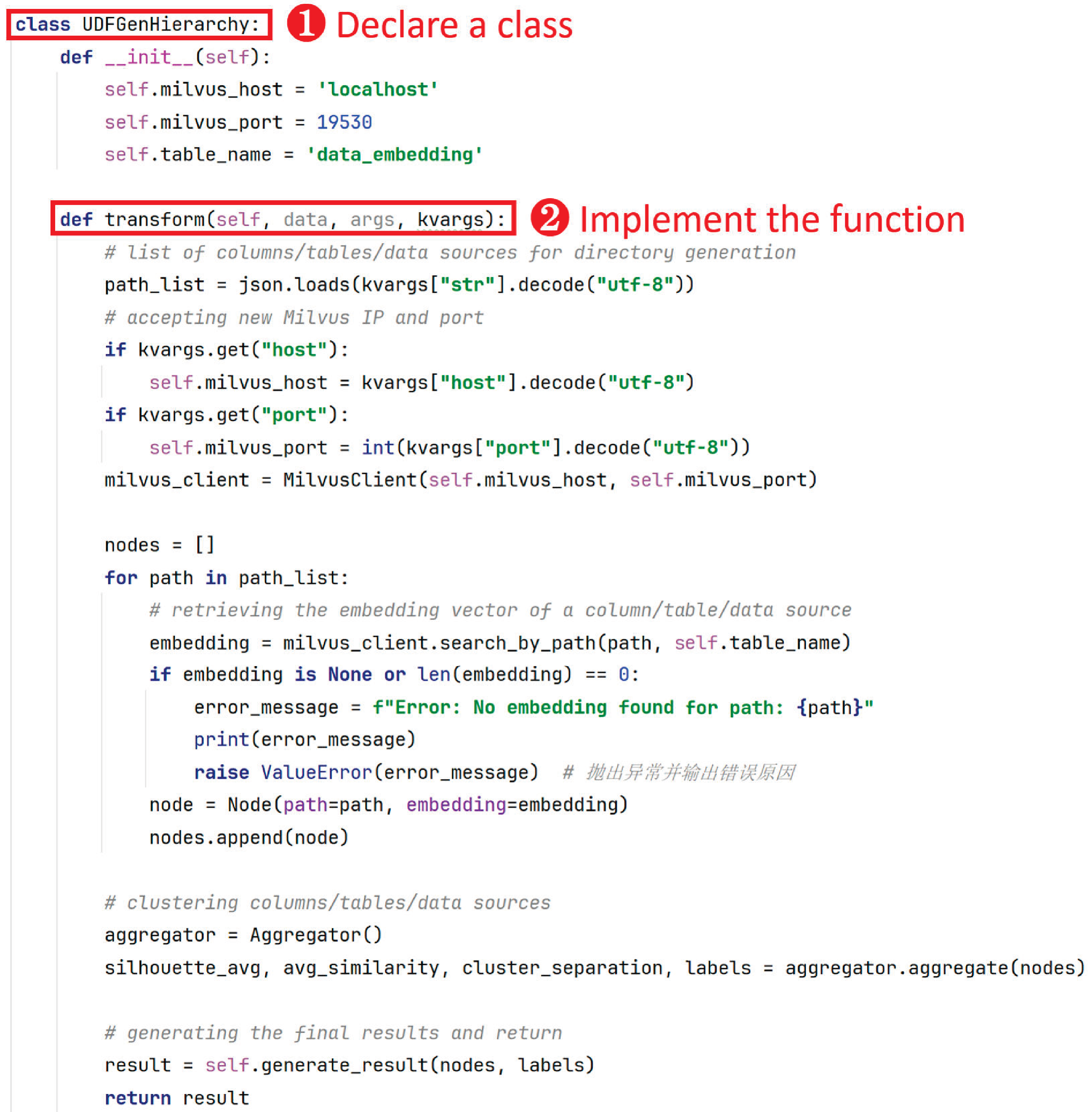}
	\caption{The Python interface and implementation for hierarchical data catalog generation.}\vspace{-12pt}
	\label{fig:hiergen}
\end{figure}

As shown in Figure~\ref{fig:flow}, the algorithms for schema clustering, query embedding, and table/category relation analysis can be easily extended or replaced by implementing a different IGinX UDF. Figure~\ref{fig:hiergen} illustrates LEDD's implementation of the IGinX UDF for generating the hierarchical data catalog. To replace the algorithm, one needs only to i) declare the same class of \emph{UDFGenHierarchy}, and ii) implement the \emph{transform} function of the class under the same signature. If one is to replace the schema clustering algorithm, one only needs to replace the \emph{Aggregator} class implementation in Figure~\ref{fig:hiergen}. Similarly, users can replace or extend the algorithms for query embedding and table/category relation analysis by implementing a different \emph{transform} function for the classes \emph{UDFGenEmbedding} and \emph{UDFAnalyzeRelation}.\vspace{-6pt}
\begin{figure*}[t]
	\centering
	\includegraphics[width=\linewidth]{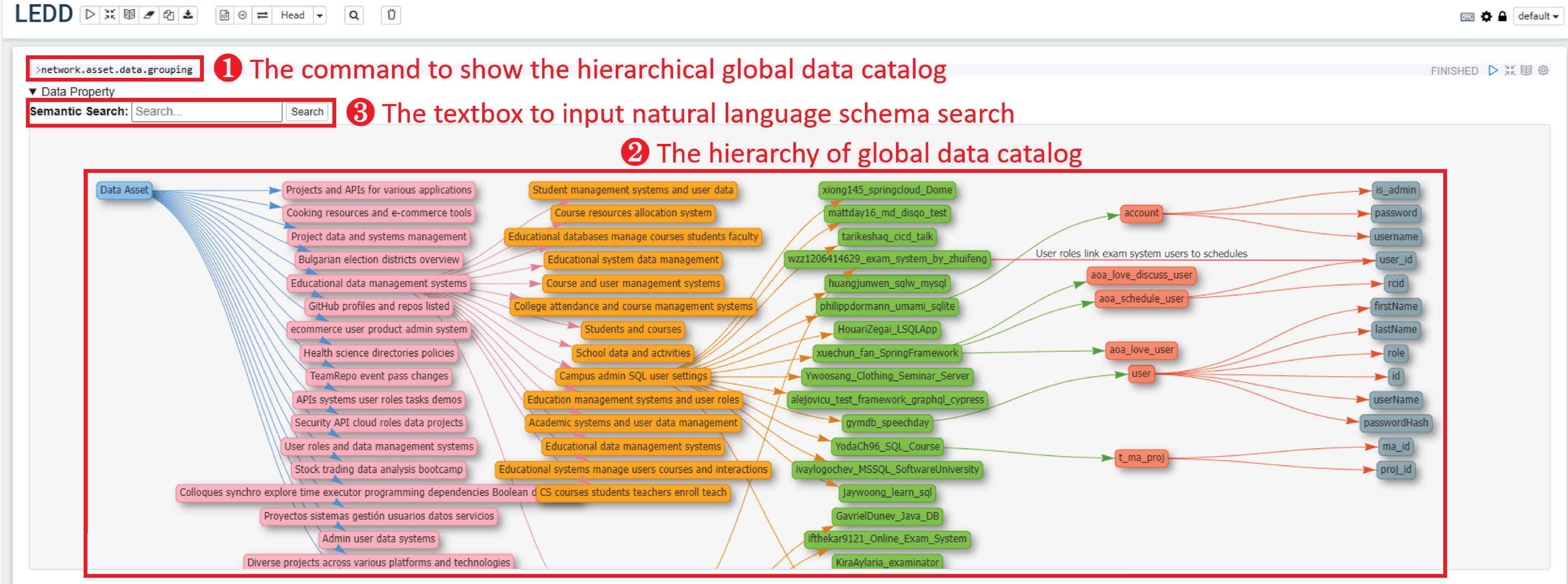}\vspace{-12pt}
	\caption{LEDD displaying the hierarchical catalog view of tables on Github.}\vspace{-12pt}
	\label{fig:hierarchy}
\end{figure*}
\section{Demonstration Scenarios}

This section showcases our demonstration scenarios, which highlight the main functionalities of LEDD. These scenarios illustrate how LEDD organizes tables within the data lake into a semantic hierarchy of data catalogs. Users can utilize LEDD functionalities via Web interfaces or IGinX's Python APIs.

\textbf{Datasets.} We exploit a recent benchmark for database schema analysis~\cite{schemapile}, involving 221,171 database schemas that are extracted from SQL files on Github. The total number of columns in the schema is about 10 million, on which we implemented and evaluated the hierarchy generation algorithms based on multiple open LLMs. Moreover, we also provided a number of typical join/union table search algorithms implemented as IGinX UDFs that can be changed or replaced flexibly.

\textbf{Hierarchical global data catalog.} LEDD can provide a hierarchical global catalog for all data in a data lake like IGinX, to facilitate data exploration and understanding. To be specific, as shown in Figure~\ref{fig:hierarchy}, the user could first see a limited list of topics that are summaries of the underlying 1.7 million tables, which are clustered and summarized into a hierarchy of topics. By clicking on any node in the graph, the user can see the corresponding topic expanded into a further list of phrases, which will finally get connected to a list of columns.

\textbf{Semantic search of tables.} LEDD allows users to input a query for tables or columns in natural language. To be specific, as shown on the top left of Figure~\ref{fig:hierarchy}, the user can fill in the textbox a keyword (e.g., "user") or a natural-language query (e.g., "info about users") to retrieve semantically relevant tables and columns. Then, tables related to user information are retrieved from IGinX and highlighted in the graph.

\textbf{Real-time relation analysis.} We will allow the user to click on the displayed graph of LEDD. In real time, the clicked node will be collapsed or expanded. If expanded, the newly displayed nodes will be analyzed by LEDD towards the original nodes for important relations, which are then represented as an edge with the relation summarized by the backend LLM. One such edge between nodes across non-neighboring layers is illustrated in Figure~\ref{fig:hierarchy}. This helps users identify potentially related datasets.

\textbf{Algorithm extension.} We will also allow users to implement their own algorithms for data discovery, as well as to choose algorithms from LEDD's library. LEDD allows users to declare new algorithms through the Web interface, using simple syntax such as "\emph{CREATE FUNCTION UDSF 'gencatalog' FROM 'UDFGenHierarchy' IN 'UDFGenHierarchy.py'}" for the class in Figure~\ref{fig:hiergen}.\vspace{-6pt}

\section{Conclusion}

In this paper, we present LEDD, an end-to-end system with an extensible architecture that exploits LLMs to provide hierarchical global catalogs with semantic meanings and semantic table search for data lakes. It supports real-time analysis of relations among categories, data sources, tables, and columns. While it supports schema clustering, query embedding, and table/category relation analysis natively, LEDD facilitates users in implementing their own algorithms for the same purposes based on a Python class interface. LEDD provides a comprehensive solution to empowering data discovery with LLM within a data lake system, with a user-friendly Web interface for flexible user interaction and downstream data applications such as training data selection or data governance.\vspace{-6pt}

%\begin{acks}
%We would like to express our sincere thanks to all those alumni that have contributed to IGinX, especially Yanzhe An, Yuan Zi, Yili Wang, Zhilin Jiang, Haozhe Li, Yu Feng, and Linzhe Zhang.
%\end{acks}

\bibliographystyle{ACM-Reference-Format}
\bibliography{ref}

\end{document}